\documentstyle[12pt,epsf]{article}
\begin{document}

\title{POLARIZATION AND COVALENCY IN ALKALI--TETRALIDE CLUSTERS.\\
}
\author{W. GEERTSMA, \\
Departamento de Fisica, UFES, Av. Fernando
Ferrari s/n\\
Vitoria--ES, Brasil.}
\maketitle

\begin{abstract}
In  this paper  we  discuss the structure  of A$_4$M$_4$  alkali(=
A)--tetralide (= M) (= group 14) clusters.   Without  polarization
these  polyions consist of  a central  tetralide tetrahedron  with
each  face capped    by  an alkali   ion.    We  show   that ionic
polarization can lead to quite different cluster   structures   by
breaking up of the tetralide tetrahedron into pairs M$_2$,  and it
can even destroy covalent  bonds  in these clusters.  Consequences
for the  structure of the solid, liquid  and amorphous phase
are discussed.
\end{abstract}

%************************************************************************

\vfill

\newpage %
\section{INTRODUCTION.}\label{section: introduction}
In this paper  we present     results of model calculations on the
atomic   structure  of A$_4$M$_4$ clusters,  where A is an alkali:
Li, Na, K, Rb, Cs, and M  a group 14 atom, also called tetralides:
Si, Ge, Sn, Pb.  We use a simple   H\"{u}ckel--type  approximation
for  the electronic  structure  and   an ionic  model      for the
interionic Coulomb and polarization interactions.

A study of  the  atomic  structure of this type of clusters is  of
importance  for the understanding of the structural and electronic
properties    of  these  clusters,  but  also  for  the associated
crystalline,  amorphous and  liquid phase. The corresponding solid
equiatomic AM compounds can roughly  be divided --  based on their
crystallographic structure  -- into three  groups (for a review of
the  structure see~\cite{review Zintl  solid 1,review  Zintl solid
2}). A group which  contains a clear three dimensional  network of
three--fold  coordination  on the  M sublattice:   LiSi  and LiGe.
Another group which contains  charged covalently   bonded    M$_4$
tetrahedra. Also in this configuration       M  has a  three--fold
coordination.    And finally the group  in which one suspects only
weak  M--M  bonds,   LiSn and  LiPb.  There   are other structures
possible  for  the   tetralide    sublattice:  for example in some
alkaline--earth  ditetralides  one finds layer structures. We will
not discuss these compounds in the present paper.

The  basis for an interpretation     of the  structure    of these
compounds  is that in these compounds one electron is  transferred
from the alkali  to the tetralide.  The local structure   of the M
sublattice of the first two groups   can be rationalized using the
Zintl \cite{Zintl}  concept:  because  one electron is transferred
from   the alkali   to the  tetralide,  the   valence     electron
configuration  of  the  M$^{-}$ tetralide    ion  of group   14 is
equivalent to that of group   15: the pnictides: P, As, Sb and Bi.
This  M$^{-}$ ion acts as a  pseudo element  with chemical bonding
characteristics equal to those of the  next group.  This is indeed
what one observes.

We have explained the difference between the compounds   with  and
without tetrahedra as due to the  size  of  the alkali \cite{paper
I} (to be referred  to  as paper I). When the alkali   ion becomes
too small  it is not able to separate the tetrahedra and the bonds
within  and between tetrahedra become the same.  In that situation
the system  may  choose   another structure for the  M sublattice,
leading  -- in our particular   case -- to the other two different
structure types   observed   for the LiM compounds.     The  large
difference between  the structures of LiSi  and LiGe on   the  one
hand and the    LiSn  and LiPb   structures  on  the other hand is
probably  due to  a difference  in covalent M--M bonding, which is
much larger  in the first  than in the second case.  We have tried
to  quantify     this but only     succeeded   in    a qualitative
way~\cite{paper III}.

In  I we  already   noted the importance      of  polarization  in
determining  the local  structure     in  these  type of  systems.
However polarization  is difficult  to take  into  account  in the
models    we used.  In order to calculate the contribution  of the
electronic  structure   to  the total energy, we approximated  the
atomic structure  of the  tetralide sublattice  by  some  type  of
pseudolattice     in which  angles   and  next--nearest  neighbour
distances  are not   well defined. Therefore we decided  to  study
smaller units,  which  are  a representation of the solid: neutral
A$_{4}$M$_{4}$   units.     Such units      were   postulated   by
\cite{Saboungi neutrons}   to explain  neutron diffraction data on
liquid  alloys  of   these compounds. Saboungi  has observed  that
these  units  seem  to rotate rather free at high temperatures  in
the solid phase without going  into a liquid phase: the so--called
rotor phase~\cite{rotors}.    There  are many more indications for
the presence of  such rather stable  entities in the liquid,  from
thermodynamic   data~\cite{specific  heat}, from  the    amorphous
phase~\cite{amorphous},    and    from  an  analysis   of  neutron
diffraction experiments\cite{RMC 1,RMC 2}.  For a recent review of
the field see for example~\cite{review 1,review 2}.

In recent  molecular dynamics simulations   of the liquid phase of
these systems, using the  Car-Parrinello   method~\cite{CP   1} --
\cite{MD  3}, one finds all kind of aggregates  on  the  tetralide
sublattice, pairs, broken  tetrahedra,   chains, and mixtures   of
chains and tetrahedra.  In an analysis one finds  however  a clear
indication for preferred  three--fold coordination    with in many
cases a peak around 60$^{0}$ in  the three--particle   correlation
function of the tetralide partial structure factor.

Recently, theoretical    studies of the structure and stability of
these   tetralide  clusters  have been  published~\cite{Valladolid
group} using the local density approximation.   These studies done
by the  Valladolid  group are most pertinent to our study, however
they neglect the core polarization, which we will  show   is  very
important in determining the atomic structure of these systems.
\section{THEORY.}
We use  a rather  simple model  to  calculate  and   minimize  the
ground state energy  with respect  to the interatomic distances and
to determine the stable atomic  structures of these clusters.   We
use a H\"uckel  type  tight-binding  model,  using Harrison's  new
parametrization   scheme~\cite{SST,Harrison          new},    with
nonorthogonal  orbitals  to calculate the electronic structure  of
these clusters.    On the alkali as well as the tetralide we  only
take into account    valence $s$ and $p$ orbitals.  The so--called
peripheral  $s$ state  correction in this new scheme is taken into
account as a perturbation.

To  this we add  the Born  repulsion   between two  atoms a distance $R_{12}$
apart, which we approximate by
\begin{equation}
F(\rho_1+ \rho_2)\exp((R^0_1 + R^0_2 -R_{12})/(\rho_1 + \rho_2)).
\end{equation}
The  Born radius $R_i^0$ of  atom   $i$   is fitted    so that the
experimental  and calculated  interatomic  distances   agree.  The
range of the   Born  repulsion    of  atom     $i$ is   taken   as
$\rho_i=R_i^0/18.6$.  We tried other   values,  but this  did  not
have a large effect on our results.     We set the prefactor  $F =
0.5e^2$, where $e$ is the electronic charge.

The van der Waals interaction is approximated by $C/R^6$, where we took the
van der Waals coefficient $C=\alpha$, i.e. equal to the polarization of the
ion.

The ionic Coulomb  interactions are calculated with the full ionic
charges. For   ionic  solids this is a good  approximation for the
lattice energy.  Whether this  approximation  also holds for these
clusters,    which   are   partially  ionic, is questionable.  The
dipolar    energy  includes     the dipole--electric        field,
dipole--dipole   and dipole self  energy.    The latter  should be
taken  into account  because   all dipoles  are   induced.     The
(maximum) polarizabilities are taken from  Fraga~\cite{Fraga}, and
are  denoted   by $\alpha  _{F}$. In  order    to   calculate  the
polarization energy  one   has  to   minimize   the  contributions
involving the induced dipole moments.  This leads to a finite  set
of linear equations in the induced dipole moment.

In order to find the minimum energy   for the atomic configuration
of a cluster,   we use a simple  simplex   scheme.   The  starting
configuration of each system is some random   configuration.  From
the electronic  structure calculation we  find a  nearly  complete
transfer of one electron from the  alkali to   the tetralide.  For
large polarizabilities   ($\alpha  >0.7\alpha   _{F}$)  the system
often becomes frozen into some  state with one very short AM bond,
leading to a large  induced dipole moment on the M ion. The length
of the dipole  moment -- based on one electron --  far exceeds the
diameter  of the atom  or ion.  In this case the   dipolar  energy
exceeds  all other  energies.  The  largest polarizability  we can
take for the tetralide without encountering  this problem is about
0.75 to 0.8 times  the Fraga values. For larger values the cluster
always finds  a configuration     with  a very  large polarization
energy.  We   also performed calculations using   the Fraga values
and varying the ionic charges.    We found that only ionic charges
up to about 0.8 give  stable clusters. For larger ionic charges we
encounter    the same problems  as reported above for the case  of
large polarizabilities.
\section{RESULTS AND DISCUSSION.}
In order to test the reliability   of this scheme for our clusters
we  calculated the  binding energy   and  vibration  energy of the
neutral    M$_{2}$  and  M$_{4}$,  and the  charged   M$_{4}^{4-}$
clusters.   In  the case  of  the neutral clusters    we  have  to
consider only  the  electronic hybridization energy  and the  Born
repulsion.    The   Born  radius   is  fitted  to the  interatomic
distances  observed   in   their crystalline structure, using  the
charged M$_{4}^{4-}$  clusters.  Using this  Born radius we obtain
good agreement  with  experiment for  the   interatomic  distance,
binding energy, and vibration  frequency  of the neutral  clusters
M$_{2}$ and M$_{4}$.  These  results  will be published  elsewhere
together with a more  detailed account of the model~\cite{geertsma
2002b}.  So we are  rather confident that the covalent  bonding is
rather  well    described    within  this model, at least for  the
interatomic distances  for these clusters.   We have applied  this
model also to other neutral clusters like MX$_{4}$,  where X is  a
halide,   and to clusters of pnictides.  We find good    agreement
with experiment   (interatomic    distances,   binding energy  and
vibration frequency) whenever available.  Results on these neutral
systems will be published elsewhere.

We varied the polarizability from zero to approximately  0.7 times
the Fraga  values. The results of these calculations are displayed
in  table  1. We find the  following    three     stable    atomic
configurations for these clusters  using  this model approach: The
Normal Double  Tetrahedron  (NDT), which consists    of  a central
M$_{4}$ tetrahedron with  the four A ions capped  outside   on its
four faces. The Face Centered Tetrahedron  (FCT), where the four A
ions are approximately near the  center   of the four faces of the
M$_{4}$ tetrahedron.  And a  configuration consisting of two pairs
of M$_{2}$ clusters,  bridged by  the four A ions. We  found other
atomic configurations  in which the simplex  got  stuck,   like  a
M$_{4}$ square,  with two A ions on  each side,  a double pyramid,
with three  short and three long bonds of the  M$_{4}$ subcluster,
and an A ion in the  base plane  (three  long bonds).  The binding
energy of  the  latter  configurations    is  some eV  above   the
ground state configuration.

We find that   without polarization   all systems     have   a NDT
ground state.  Molina et al~\cite{Valladolid   group} find the same
atomic    configuration for Li$_{4}$Pb$_{4}$ and  Na$_{4}$Pb$_{4}$
clusters. Turning on the polarization  however causes a separation
of   these  clusters   in three  groups:      The  group where the
ground state remains a NDT  configuration up to 0.6 times the Fraga
polarizability:        these   are     all   the   K$_{4}$M$_{4}$,
Rb$_{4}$M$_{4}$,  and Cs$_{4}$M$_{4}$clusters,   the  group  where
there is a transition   to the FCT state,  without    -- or nearly
without   -- an  intermediate    state  with M$_{2}$ pairs in  the
Li$_{4}$Sn$_{4}$,     Li$_{4}$Pb$_{4}$,    Na$_{4}$Sn$_{4}$    and
Na$_{4}$Pb$_{4}$ clusters, and the group   where  there is a clear
intermediate  state   with pairs, before   the  cations   move  to
approximately the center  of the  faces of the M$_{4}$ tetrahedron
(Li$_{4}$Si$_{4}$,          Li$_{4}$Ge$_{4}$,    Na$_{4}$Si$_{4}$,
Na$_{4}$Ge$_{4}$).   In figure 1 we  illustrate these   structures
for the   Li$_{4}$Si$_{4}$ cluster: for small polarizability   the
NDT structure     has  the lowest    energy,    for   intermediate
polarizability the structure   with  two  Si$_{2} $  pairs is most
stable, and for large polarizability of  Si the FCT structure  has
lowest energy.

Let us next   try to explain these structural  transitions.   When
the polarization  is  turned on,  the cluster  can  and  wants  to
increase  the  polarization  contribution to  the  total   energy,
however  this will cost Coulomb   as well as hybridization energy.
When the  alkali ions are outside  the M$_{4}$    tetrahedron, the
electric field on the M ion is rather small, as the fields  of the
alkali and  the  tetralide nearly  cancel.    The  way   to  gain
polarization energy is to move the  alkalis    from far outside to
more near  the faces of the M$_{4}$ tetrahedra,  however this will
cause an increase  of  the interatomic   distances  of the M$_{4}$
tetrahedron.  When the covalent bonding is large  like in the case
of Si and Ge, an intermediate  state can be created,  in which the
Si  or  Ge form  pairs, with a relative  large covalent    bonding
energy. Further  increase of the polarization  causes also a break
up of these    M$_{2}$    pairs,    and  we   finally  have  a FCT
configuration,   where  the covalent bonding between the M ions is
weakened,  although in the  case of Si and Ge  it still gives   a
relative large contribution to the total binding energy.

We note  that  when we increase the polarization   the  M--M  bond
length  in general decreases, except in the case Li$_{4}$Pb$_{4}$,
Li$_{4}$Sn$_{4}$,  Na$_{4}$Pb$_{4}$.    This decrease is caused by
the fact    that the  electric field   increases  with  decreasing
bondlenght,     thereby    increasing    the polarization   energy
contribution   to the binding energy. The M--M  bondlenght for two
values ($\alpha =0$,  and $.5\alpha  _{F}$)  of the polarizability
are given in table 1. The increase of the M--M distance when   the
polarization      is  turned    on   is the precursor for the  NDT
$\rightarrow $ FCT transition.

From our   model calculations   we find rather natural   the often
observed tetrahedral   atomic configuration for Li  and less often
observed  for   Na~\cite{Li       bonding 1,Li bonding  2}.   This
configuration is especially stable for counter--ions with a  large
polarizability, like Pb$^{-}$.  This atomic configuration   of the
Li$_4$ cluster  is -- in the systems studied  here --  not  due to
direct covalent bonding between the Li ions,  but due polarization
effects. The  Li $p$ levels are too  high in energy to participate
in bonding.  The  M$_{4}$  have become essentially noninteracting,
this  subcluster   is in  this   limit (FCT) not stable on its own
account.

Finally, in figure 2 we  present results on the calculation of the
atomic  structure of Cs$_{n}$Pb$_{4}$    clusters for $n=1$, 2, 3.
For  $n=1$ we find a capped tetrahedron, for $n=2$ a square of Pb,
with  a Cs  on  each square face,  and  for  $n=3$    we   find  a
three--capped   tetrahedron.  Such clusters have also been studied
by the Valladolid group~\cite{Valladolid   group} for  Li--Pb  and
Na--Pb.   Our structures differ appreciably from theirs  for $n=1$
and for $n=2$.  However   for  $n=3$ our  structure for  Cs--Pb is
very similar to their structures.   In their  calculation they did
not  take into    account the core  polarization.  Furthermore the
local  density  approximation  using pseudopotentials  is probably
less adequate to describe the electronic  and therefore the atomic
structure of clusters with strong  ionic  bonding.  It would be of
interest to study alkali--rich   clusters,  especially for Si  and
Ge.

Let us now discuss  the  observed  crystalline structures based on
the results of the present calculations.  First  we  note that the
electric fields due to the  ions in the condensed phase can  never
attain such large  values as one  finds in   a cluster.   However,
starting  with the  NDT  configuration  without polarization, also
the  condensed phase can lower its energy by changing  the bonding
in the M sublattice.   Such changes in bonding are   more   easily
accomplished in a liquid or amorphous state than in  a crystalline
solid. In the latter case periodicity imposes  restrictions on the
local  structure.   However     a  liquid  exists     only at high
temperatures   and we \cite{specific heat} have seen that at these
temperature  notwithstanding the  relative high  binding energy of
the clusters, entropic    effects    cause   these   clusters   to
dissociate.  At  very  low temperature  in the amorphous phase  we
expect these  clusters  to  exist     as  rather stable  entities,
localizing  the   valence   electrons     in their  covalent bonds
\cite{amorphous}.     This  is  actually the interpretation of the
lack  of the  peak in the resistivity for  Na$_x$Pb$_{1-x}$  as a
function   of $x$ in the liquid state:  there are no  M$_{4}^{4-}$
tetrahedra to capture  the conduction electrons in its bonds. That
is, the  NDT $\rightarrow    $ FCT  transition in NaPb  occurs  on
melting. In the amorphous phase one observes such a peak.

Systems  where such  a transition  away from the NDT configuration
is possible  with the lowest fraction  of the polarizability   are
the LiM  systems.   One   notices also that LiSi  and  LiGe have a
strong tendency  to  form pairs, while   LiSn has a small tendency
and LiPb has  a direct  transition    from  the  NDT   to  the FCT
configuration.    This  could explain  why in  the solid state the
LiSi  and LiGe  form a three-dimensional network,  while  LiSn and
LiPb do not.  So, based on polarization,  we have a   rather clear
separation   between  the LiM systems and  the  other   A$_4$M$_4$
systems and  within  the LiM systems  between LiSi and LiGe on the
one side and LiSn and LiPb on the other side.

The hardest bonds are  the M$_4$  bonds, while the  MA  bonds  are
much  weaker. So by applying  pressure  one will shorten MA bonds,
thereby increasing  the polarization  energy contribution, leading
to a break up of the M$_4$  tetrahedron. We expect that one of the
easiest systems  to transform   from  the NDT   to the   FCT--like
structure is the  NaPb system:   the electric fields,  causing the
polarization break up of the covalent MM bonds,   are   relatively
large,  while the covalent interactions  in  the M$_4$ tetrahedron
are relatively weak.  So under pressure we  expect it to transform
from a system with tetrahedra to one without. 
\section{CONCLUSIONS.}
Based on our    model  calculation   we  conclude   that the ionic
polarization cannot be neglected in the  calculation of the atomic
structure    of   polyions, with  partially  ionic  bonding,  i.e.
cluster  consisting      of atoms with   a   large  difference  in
electronegativity.   We also have  seen   that polarization energy
gain  can break up even relatively  strong covalent  bonds like in
the case of LiSi and LiGe.  Clearly, effects  due  to polarization
are  in  these cluster  opposite   to those of   covalent bonding:
covalent   bonding favors  a NDT    structure, while  polarization
favors a FCT structure.

\section*{Acknowledgments}
We acknowledge a grant from the European Union ( TMR contract:
ERBFMBICT--950218), which made the first stages of this research possible,
and a grant from CNPq (300928/97--0) during the final stages of this work.
\newpage

\begin{table}[tbp] \centering%
\caption[table     1]{\label{table  1}Stable atomic configurations
found   for  the A$_{4}$M$_{4}$  clusters,    as  a  function   of
polarizability.\newline  NDT: Double tetrahedron with  the alkalis
capped outside  on the four faces   of the tetralide  tetrahedron.
\newline FCT:  The same as NDT,   but with each alkali capped near
the center of  the  one of the four  faces.   
\newline Pairs:  Two
tetralide pairs, bridged by alkali ions.  
\newline  The   value of
the fraction of the maximum polarization  ($\alpha   _{F}$)  where
the transition   takes  place from one cluster structure  to
another is indicated  above the  arrows. The error  is about  $\pm
0.02$.  Below     the arrows we give the interatomic M--M distance
(in $\AA $)  for zero polarizability and for a finite value of the
polarizability. The polarizability is given in parenthesis. For
details see the main text.}%
\end{table}%
{\scriptsize 
\begin{tabular}{|cc|cc|cc|}
\hline
&  & Si & Ge & Sn & Pb \\ 
&  & $R^{0}=1.2$; $\alpha =7.27$ & $R^{0}=1.31$; $\alpha =7.51$ & $R^{0}=1.49
$;$\alpha =10.8$ & $R^{0}=1.62$;$\alpha =12.4$ \\ \hline
{\scriptsize Li} & $R^{0}=1.25$ & NDT $\stackrel{0.31}{\rightarrow }$\ pairs 
$\stackrel{0.56}{\rightarrow }$\ FCT & NDT $\stackrel{0.35}{\rightarrow }$\
pairs $\stackrel{0.48}{\rightarrow }$\ FCT & NDT $\stackrel{0.31}{%
\rightarrow }$\ pairs $\stackrel{0.39}{\rightarrow }$\ FCT & NDT $\stackrel{%
0.34}{\rightarrow }$\ FCT \\ 
& $\alpha =0.003$ & 2.60(0.0);2.53(0.25) & 2.71(0.0);2.66(0.25) & 
3.09(0.0);3.13(0.25) & 3.36(0.0);3.49(0.25) \\ \hline
Na & $R^{0}=1.35$ & NDT $\stackrel{0.55}{\rightarrow }$\ pairs $\stackrel{%
0.80}{\rightarrow }$\ FCT & NDT $\stackrel{0.51}{\rightarrow }$\ pairs $%
\stackrel{0.70}{\rightarrow }$\ FCT & NDT $\stackrel{0.44}{\rightarrow }$\
pairs $\stackrel{0.46}{\rightarrow }$\ FCT & NDT $\stackrel{0.37}{%
\rightarrow }$\ FCT \\ 
& $\alpha =0.155$ & 2.65(0.0);2.49(0.25) & 2.80(0.0);2.66(0.25) & 
3.15(0.0)3.48(0.25) & 3.43(0.0);3.45(0.25) \\ \hline
K & $R^{0}=1.59$ & NDT $\stackrel{>0.75}{\rightarrow }$\ pairs/FCT & NDT $%
\stackrel{>0.75?}{\rightarrow }$\ pairs/FCT & NDT $\stackrel{0.72}{%
\rightarrow }$\ pairs/FCT & NDT $\stackrel{0.60}{\rightarrow }$\ pair$%
\stackrel{0.63}{\rightarrow }$\ FCT \\ 
& $\alpha =0.947$ & 2.87(0.0);2.42(0.5) & 3.00(0.0)2.57(0.5) & 
3.34(0.0);2.98(0.5) & 3.61(0.0);3.33(0.5) \\ \hline
Rb & $R^{0}=1.71$ & NDT $\stackrel{>0.75}{\rightarrow }$\ pairs/FCT & NDT $%
\stackrel{>0.75}{\rightarrow }$\ pairs/FCT & NDT $\stackrel{>0.75}{%
\rightarrow }$\ pairs/FCT & NDT $\stackrel{0.74}{\rightarrow }$\ pairs \\ 
& $\alpha =1.65$ & {\scriptsize 3.03(0.0);2.42(0.5)} & 3.15(0.0);2.57(0.5) & 
3.48(0.0);2.9(0.53) & 3.73(0.0);3.31(0.5) \\ \hline
Cs & $R^{0}=1.82$ & {\scriptsize NDT }$\stackrel{>0.75}{\rightarrow }$%
{\scriptsize \ pairs} & NDT $\stackrel{>0.75}{\rightarrow }$\ pairs & NDT $%
\stackrel{>0.75}{\rightarrow }$\ pairs & NDT $\stackrel{0.81}{\rightarrow }$%
\ FCT \\ 
& $\alpha =3.08$ & {\scriptsize 3.19(0.0);2.41(0.5)} & 3.40(0.0);2.57(0.5) & 
3.63(0.0);2.97(0.5) & 3.87(0.0);3.30(0.5) \\ \hline
\end{tabular}
} 

\newpage 
%************************************************************************
\pagebreak

\section*{Figure Captions}
\begin{figure}[h]
\caption[fig 1]{The atomic structure of  Li$_{4}$Si$_{4}$ clusters
as a function   of the polarizability:  A  $\alpha  =0$  (NDT);  B
$0.4\alpha  _{F}$ (two  pairs);  C $ 0.55   \alpha   _{F}$  (FCT);
$\alpha  _{F}$   is  the   value     for     the    polarizability
from~\protect\cite{Fraga}. } \label{LiSi}
\end{figure}
\begin{figure}[h]
\caption[fig 2]{The atomic structure  of the Cs$_{n}$Pb$_{4}$
clusters for
n  =1, 2,  3 clusters  with polarizability:  $0.25\alpha    _{F}$;
$\alpha    _{F}$     is     the value   for  the    polarizability
from~\protect\cite{Fraga}. } \label{CsnPn4}
\end{figure}

\pagebreak
\epsfverbosetrue
\epsfysize=5in
\epsffile{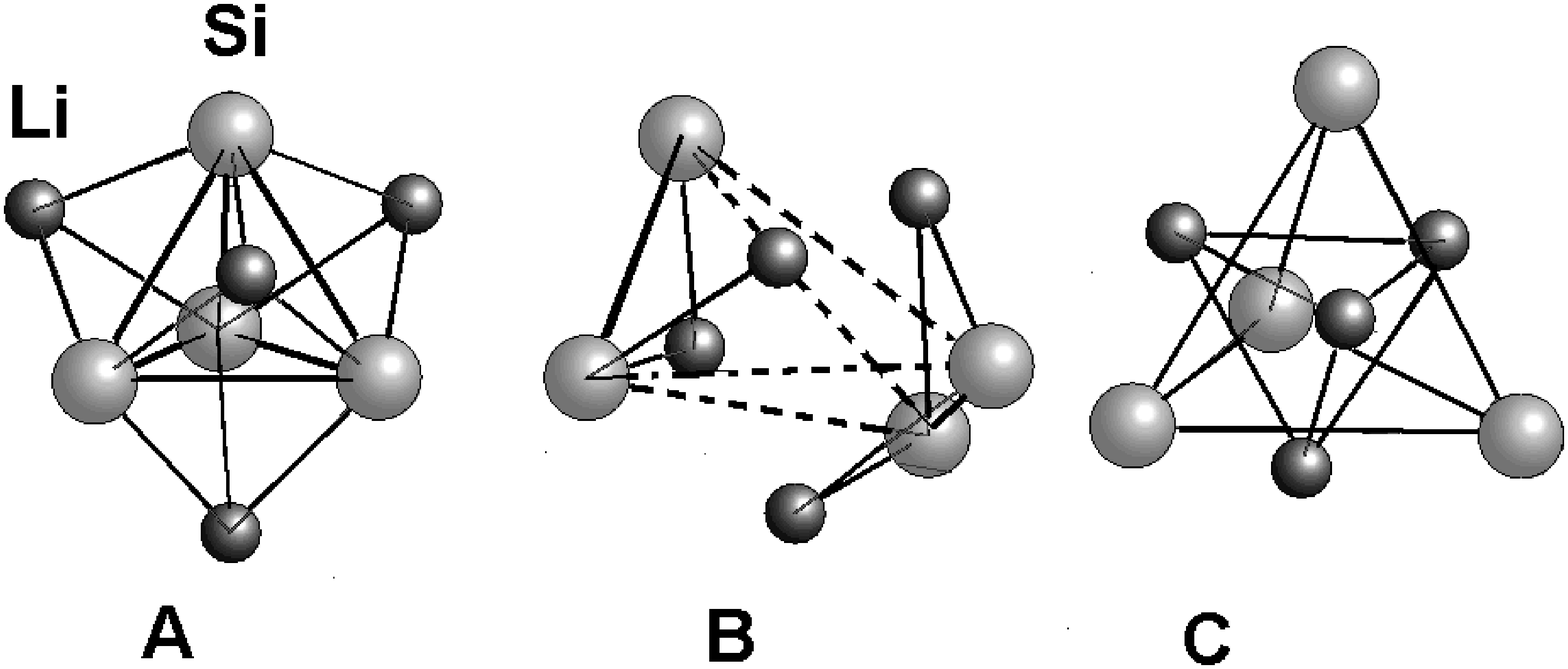}
\epsfysize=5in
\epsffile{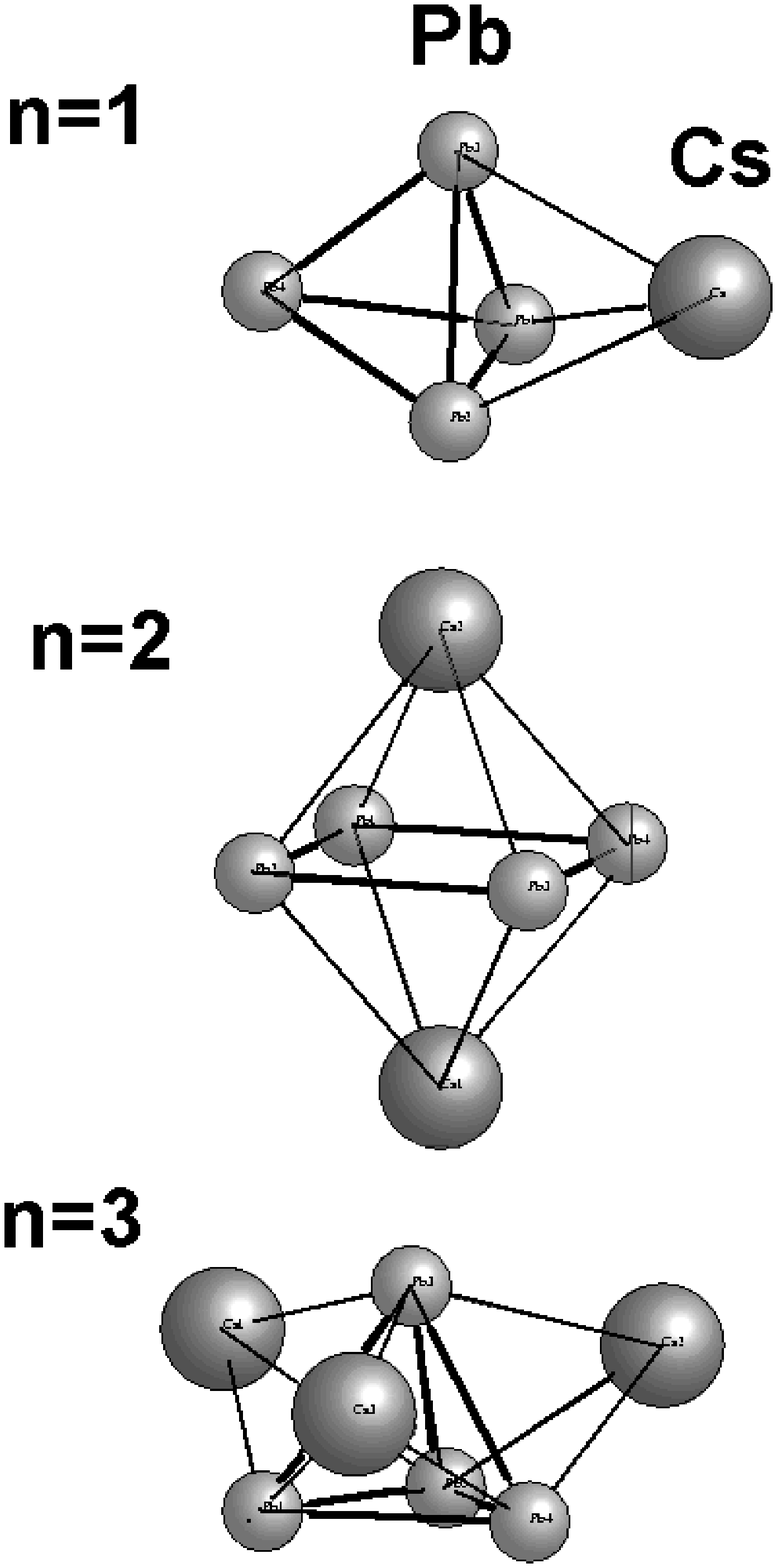}
\end{document}